\documentclass[a4paper,fleqn,usenatbib]{mnras}

\usepackage{newtxtext,newtxmath}

\usepackage[T1]{fontenc}
\usepackage{ae,aecompl}

\usepackage{graphicx}	
\usepackage{amsmath}	
\usepackage{xcolor}
\usepackage{lastpage}

\newcommand{\msun}{\mathrm{M}_\odot}

\newcommand{\eej}{E_\mathrm{ej}}
\newcommand{\esh}{E_\mathrm{sh}}
\newcommand{\ej}{E_\mathrm{j}}

\newcommand{\cmspg}{\,\mathrm{cm}^2 \mathrm{g}^{-1}}
\newcommand{\gcc}{\,\mathrm{g}\, \mathrm{cm}^{-3}}
\newcommand{\K}{\,\mathrm{K}}
\newcommand{\erg}{\,\mathrm{erg}}
\newcommand{\kb}{k_\mathrm{B}}
\newcommand{\dg}{^\circ}
\newcommand{\ye}{Y_\mathrm{e}}
\newcommand{\tj}{t_\mathrm{J}}
\newcommand{\ts}{t_\mathrm{S}}

\makeatletter
\renewcommand\d[1]{\mspace{6mu}\mathrm{d}#1\@ifnextchar\d{\mspace{-3mu}}{}}
\makeatother

\title[Jet-Ejecta Interaction and Kilonova Light Curves]{The Effect of Jet-Ejecta Interaction on the Viewing Angle Dependence of Kilonova Light Curves}

\author[H. Klion et al.]{
Hannah Klion,$^{1}$\thanks{E-mail: hklion@berkeley.edu}
Paul C. Duffell,$^{2}$
Daniel Kasen$^{1}$
and Eliot Quataert$^{1,3}$
\\
$^{1}$Physics and Astronomy Departments and Theoretical Astrophysics Center, University of California, Berkeley, CA, USA\\
$^{2}$Center for Astrophysics, Harvard University, Cambridge, MA, USA\\
$^{3}$Department of Astrophysical Sciences, Princeton University, Princeton, NJ
}

\date{Accepted XXX. Received YYY; in original form ZZZ}

\pubyear{2020}

\begin{document}
\label{firstpage}
\pagerange{\pageref{firstpage}--\pageref{lastpage}}
\maketitle

\begin{abstract}
The merger of two neutron stars produces an outflow of radioactive heavy nuclei. Within a second of merger, the central remnant is expected to also launch a relativistic jet, which shock-heats and disrupts a portion of the radioactive ejecta. Within a few hours, emission from the radioactive material gives rise to an ultraviolet, optical, and infrared transient (a kilonova). We use the endstates of a suite of 2D relativistic hydrodynamic simulations of jet-ejecta interaction as initial conditions for multi-dimensional Monte Carlo radiation transport simulations of the resulting viewing angle-dependent light curves and spectra starting at $1.5\,\mathrm{h}$ after merger. We find that on this timescale, jet shock heating does not affect the kilonova emission. However, the jet disruption to the density structure of the ejecta does change the light curves. The jet carves a channel into the otherwise spheroidal ejecta, revealing the hot, inner regions. As seen from near $(\lesssim 30\dg)$ the jet axis, the kilonova is brighter by a factor of a few and bluer. The strength of this effect depends on the jet parameters, since the light curves of more heavily disrupted ejecta are more strongly affected. The light curves and spectra are also more heavily modified in the ultraviolet than in the optical.
\end{abstract}




\section{Introduction}

The first gravitational wave detection of a binary neutron star merger (GW170817; \citet{abbott:17_nsgw}) was followed by counterparts across the electromagnetic spectrum.
Notably, the UV/optical/IR (hereafter UVOIR) counterpart, AT2017gfo, showed the ejection of $>0.01\msun$ of material with an opacity of $1-3\cmspg$ and a kinetic energy of $\sim 10^{51}\erg$ \citep{kasliwal:17, drout:17, villar:17, coulter:17,cowperthwaite:17}. 
This is consistent with theoretical predictions for a kilonova, a red, rapidly-evolving, transient predicted to be the UVOIR counterpart of a neutron star merger \citep{metzger:10, metzger:12, barnes:13}.

The material ejected from the binary neutron star system during and immediately following merger has a low electron fraction $\ye$.
During its rapid expansion from nuclear densities, the material is a prime site for rapid (r-) process nucleosynthesis \citep{lattimer:77, eichler:89, metzger:10}.
The r-process produces a wide range of heavy, unstable, neutron-rich isotopes.
As these nuclei decay back to stability, they heat the surrounding material, powering the transient \citep{li:98}.

The light curve of AT2017gfo can be modeled by two components associated with ejecta of different opacity \citep{cowperthwaite:17, villar:17}. Both components are consistent with being heated by r-process decays. At early times, emission is largely due to a ``blue'' component associated with material rich in elements near the first r-process peak ($A \sim 80$). Within a few days, there is a transition to the ``red" emission, from material enriched by high-opacity Lanthanides around the second r-process peak ($A \sim 140$). 

GW170817 was accompanied by a low-energy short gamma ray burst $\sim 1.7 \, \mathrm{s}$ after the gravitational wave detection \citep{abbott:17_grgw}.
While the exact gamma ray emission mechanism remains uncertain, subsequent radio observations have found that the afterglow appears to move superluminally across the sky and is consistent with viewing a collimated jet at {$20\dg$} off axis \citep{mooley:18_superlum}. 

A successful collimated jet, by definition, has tunneled through the surrounding ejecta.
The jet propagation will undoubtedly affect the structure of the slower radioactive ejecta ($v \sim 0.1c$), potentially leaving an imprint on the kilonova itself.

Most studies of jet effects on kilonova light curves have focused on the effect of shock heating on the distribution of thermal energy.
Numerical simulations show that the shock-heated cocoon surrounding a jet may dominate the light curve in the first $\sim \mathrm{hour}$ after merger \citep{kasliwal:17, gottlieb:18}.
\citet{piro:18} argue that the blue component of the kilonova emission could be entirely due to jet shock heating, as opposed to the radioactive decay of r-process elements. As discussed later in this paper, we find this unlikely.

Advanced LIGO is expected to detect around 0.1 to 3 neutron star mergers per year, some of which will have detectable electromagnetic counterparts \citep{abbott:19}.
Our viewing angle relative to the binary plane for each of these events will be different.
Accurate interpretation of these events will rely on understanding how identical events would appear to different observers. This is in addition to the expected diversity in the observations due to intrinsic differences in the amount, composition, and distribution of the ejecta \citep{gompertz:18}.
Further, even if all neutron star mergers launch jets, engine energy, duration, and opening angle may all vary as well. {It is expected that some components of neutron star merger ejecta are non-spherical, and may show latitudinal variation in density and opacity.} Recent work has studied how ejecta components with different  geometry and opacity can affect kilonova light curves \citep{wollaeger:18, kawaguchi:20, darbha:20, korobkin:20, nativi:21}. In this study, we focus on {another aspect of the asymmetry:} the consequences of a jet evacuating a narrow cavity within the bulk of the ejecta. We also assess whether jet shock heating contributes to kilonova light curves on timescales longer than an hour.

 We previously performed a suite of 2D relativistic hydrodynamic simulations of a jet interacting with a homologously expanding outflow \citep{duffell:18}. In these models, we varied jet energy and opening angle, but did not account for the possible delay between the launch of the wind and the start of the jet. As detailed in section \ref{sec:simulations}, we select a representative sample of these models and calculate viewing-angle dependent kilonova light curves using Sedona, a Monte Carlo radiation transport code. We present and discuss our results in section \ref{sec:results}. We find that by 1 hour after merger, the thermal energy deposited by the jet is negligible relative to the energy generated by r-process decays. The jet, however, substantially alters the density distribution of the ejecta near the pole. The channel carved by the jet can lead to brighter and bluer emission along the axis of the jet. We conclude and discuss observational implications in section \ref{sec:discussion}.

\section{Radiation Transport Simulations}
\label{sec:simulations}
\subsection{Initial Models}
\label{sec:initial_models}
\begin{figure}
  \includegraphics{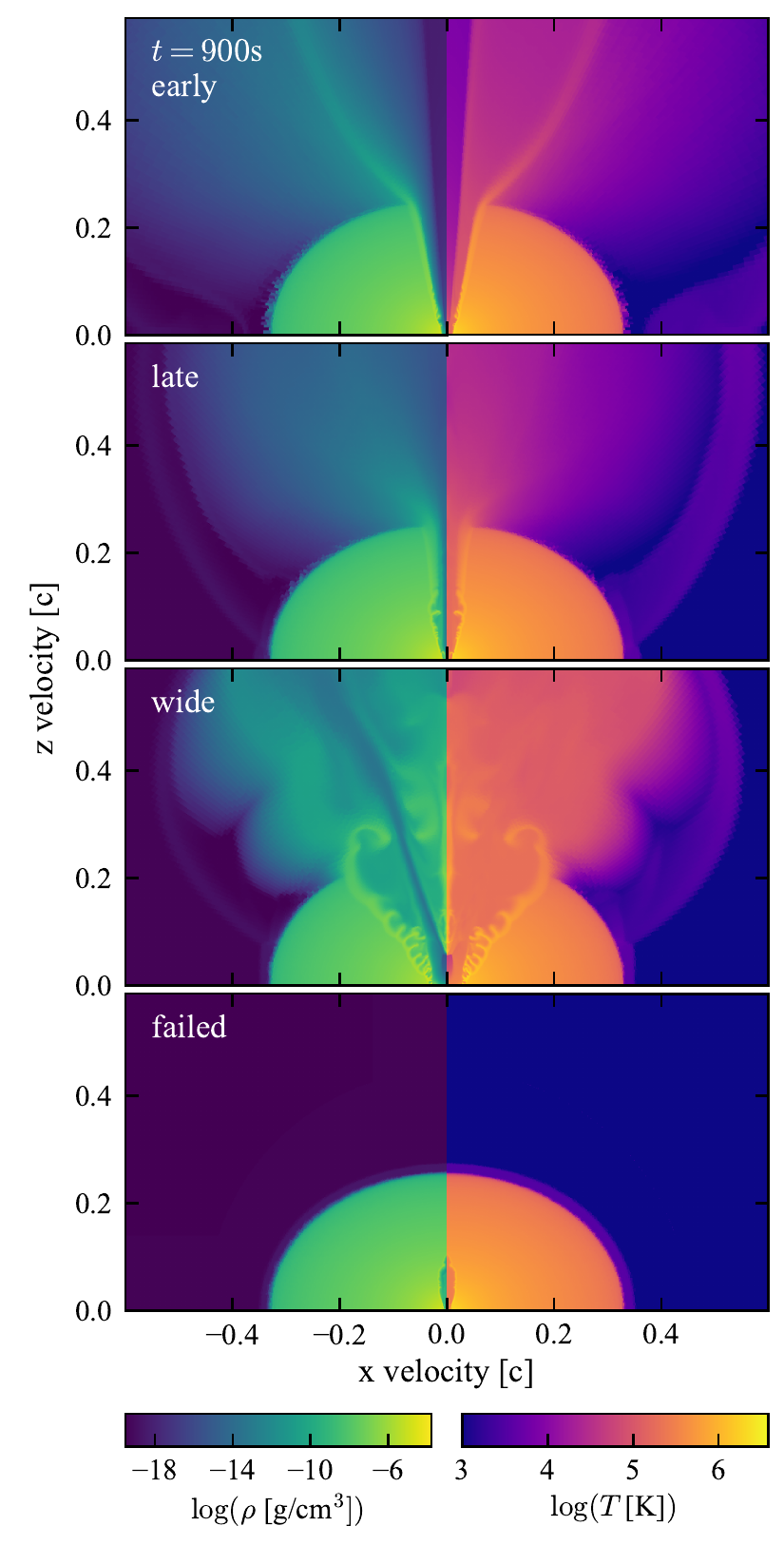}
  \caption{Starting conditions for our simulations, at 900 seconds (15 minutes) after homologous expansion and jet launching begin. The left and right panels show the logarithms of the mass density $\rho$ and temperature $T$, respectively. Each set of panels corresponds to a different type of outcome from the jet-ejecta interaction, as identified in \citet{duffell:18} and described in Table \ref{tab:models}.}
    \label{fig:2d_rho_T}
\end{figure}

In \citet{duffell:18} we studied the dynamics of a jet interacting with a homologously expanding outflow, varying jet energy and opening angle.
Our calculations were in the limit where the delay between the launch of the outflow and the start of the jet engine is very small {($10^{-4}\,\mathrm{s}$). This is much shorter than both the engine duration ($0.1\,\mathrm{s}$) and the time taken for the jet to break out of the ejecta ($3\times 10^{-3} \,\mathrm{s}$ in the fastest case).}
We performed 2D axisymmetric relativistic hydrodynamic simulations using JET, a moving mesh hydrodynamics code \citep{duffell:11, duffell:13}.
We identified the conditions under which jets can break out of the surrounding ejecta, as well as scaling relations for the time of breakout and amount of energy thermalized.
The initial structure of the ejecta follows numerical results of dynamical ejecta from neutron star merger simulations \citep{hotokezaka:13_massej, nagakura:14} {and is given in equations 15 through 19 of \citet{duffell:18}}. This component is expected to make up a small portion ($\sim 10^{-3}\msun$) of the total ejecta. However, for the sake of simplicity we assume that all of the ejected mass ($>10^{-2}\msun$) follows this structure. This is a crude approximation of NS merger ejecta, but our focus here is on the qualitative viewing-angle effects of a jet interacting with a radioactive outflow.

\begin{table}
	\centering
  \caption{Jet injection energies ($\ej$) and half-opening angles ($\theta_{\mathrm{j}}$) for the \citet{duffell:18} models that we study in this paper. The short names will be used to refer to these models throughout the paper. The models are scaled so the total mass is $4\times 10^{-2}\msun$, and the jet is active for $0.1 \,\mathrm{s}$. The jet energy is given both in ergs and as a fraction of the ejecta kinetic energy $\eej = 5.7 \times 10^{50}$.}
	\label{tab:models}
	\begin{tabular}{lcccr} 
		\hline
		Model & Short Name & $\ej$ & $\ej/\eej$ & $\theta_\mathrm{j}$\\
		& & [erg] & & [rad]\\
		\hline
		early jet breakout & early & $3\times 10^{49}$ & $5\times10^{-2}$ & 0.1\\
		late jet breakout & late & $5 \times 10^{47}$ & $8 \times 10^{-4}$ & 0.1\\
		failed jet, breakout & wide & $1 \times 10^{50}$ & $3 \times 10^{-1}$ & 0.4\\
		failed jet, no breakout & failed & $1 \times 10^{47}$ & $2 \times 10^{-4}$ & 0.1\\
		\hline
	\end{tabular}
\end{table}

Depending on the opening angle and jet energy relative to ejecta mass, we identify four qualitatively different outcomes that can arise from a jet interacting with a homologously expanding outflow:
\begin{itemize}
    \item early jet breakout (hereafter ``early''): jet breaks out before the central engine turns off.
    \item late jet breakout (``late''): jet stays collimated but breaks out after the central engine turns off.
    \item failed jet, successful breakout (``wide''): energetic jet that fails due to its wide opening angle. The large amount of deposited energy still leads to a shock breakout, though not of a jet.
    \item failed jet, no breakout (``failed''): jet completely fails and there is no breakout.
\end{itemize}

In this paper, we investigate the effect of the central engine on the kilonova light curves in each of these four cases. We choose one model from each category of outcome. Model parameters are shown in Table \ref{tab:models}. We scale the hydrodynamic calculations such that the jet duration is $0.1\, \mathrm{s}$ and the ejecta mass is $0.04 \msun$. This corresponds to an ejecta kinetic energy $\eej = 5.7 \times 10^{50} \erg$. 

The models from \citet{duffell:18} have been evolved in JET for $\tj \equiv 100\, \mathrm{s}$, until they are approximately homologous (i.e. $v \propto r$). After the JET calculation, we transform the models onto an axisymmetric velocity grid, and exclude all material with $v>0.6$. There is a negligible amount of mass at these high velocities, so it will not affect the light curves on timescales of an hour or more. The sole source of thermal energy in the JET models is the jet shocking and heating the surrounding ejecta. We denote the total of this thermal energy $\esh$, and the corresponding energy density $\varepsilon_\mathrm{sh}$. At $\tj$, the optical depth of the material is very high, so starting the Monte Carlo radiation transport calculation at that time would be computationally infeasible. Fortunately, doing so is unnecessary. Assuming homology and adiabatic expansion, we can calculate the mass density ($\rho$) and temperature ($T$) structure of the ejecta when the Sedona calculation starts at $\ts \equiv 900\,\mathrm{s}$. We can relate $\rho$ and  $\varepsilon_\mathrm{sh}$ at $\tj$ and $\ts$ by
\begin{equation}
  \label{eq:homol_rho}
\rho(\ts, v_r, \theta) = \rho(\tj, v_r, \theta) \left(\frac{\tj}{\ts}\right)^3
\end{equation}
and
\begin{equation}
  \label{eq:homol_eps}
  \varepsilon_\mathrm{sh}(\ts, v_r, \theta) = \varepsilon_\mathrm{sh}(\tj, v_r, \theta) \left(\frac{\tj}{\ts}\right)^4,
\end{equation}
respectively, where $v_r$ is the radial velocity coordinate, and $\theta$ is the latitude. Equation \ref{eq:homol_eps} accounts for the $t^{-1}$ decline in total thermal energy due to adiabatic expansion.

The total thermal energy density at a point at $\ts$ is the sum of the components from shock heating (equation \ref{eq:homol_eps}) and from radioactive heating.
The JET calculations do not account for the radioactive heating, so we add it in when constructing the Sedona models.
For a radioactive heating rate per unit mass $\eta(t)$, a fluid element of mass $m$ will have a thermal energy due to radioactive heating, $e_\mathrm{rad}$ that evolves according to
\begin{equation}
    \label{eq:rad_therm_diff}
    \frac{\d e_\mathrm{rad}}{\d t} = \frac{-e_\mathrm{rad}}{t} + \eta(t) m.
\end{equation}
Therefore, the energy density due to r-process heating at a given point in space will be
\begin{equation}
    \label{eq:rad_therm_int}
    \varepsilon_\mathrm{rad}(\ts, v_r, \theta) = \frac{\rho(\ts, v_r, \theta)}{t} \int_{t_0}^{\ts} \eta(t) t \d t.
\end{equation}
The total thermal energy density at a given point at $t_S$ is the sum of the shock and r-process contributions:
\begin{equation}
    \label{eq:homol_total_therm}
    \varepsilon(\ts, v_r, \theta) \equiv \varepsilon_\mathrm{sh}(\ts, v_r, \theta) + \varepsilon_\mathrm{rad}(\ts, v_r, \theta).
\end{equation}

\begin{figure}
    \centering
    \includegraphics{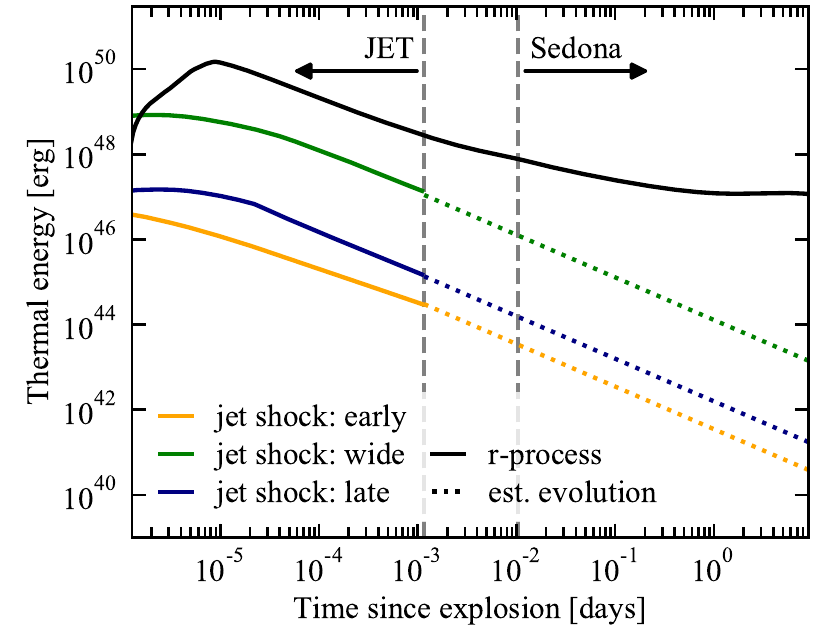}
    \caption{Estimated thermal energy of ejecta due to r-process heating (black, computed according to equation \ref{eq:rad_therm_int}), and due to jet shock heating.
    Different colors correspond to our different models (orange: early, green: wide, blue: late).
    The dashed grey lines delimit the two phases of the calculation: the hydrodynamic simulations in JET (before $100\, \mathrm{s} \approx 10^{-3}\,\mathrm{days}$), and the Sedona calculations that begin at $900\,\mathrm{s} \approx 10^{-2}\,\mathrm{days}$.
    The solid colored lines show the thermal energy in the JET simulations. The dotted lines show how the thermal energy would evolve while the ejecta are expanding adaiabatically.}
    \label{fig:rad_heating}
\end{figure}

\begin{figure}
  \includegraphics{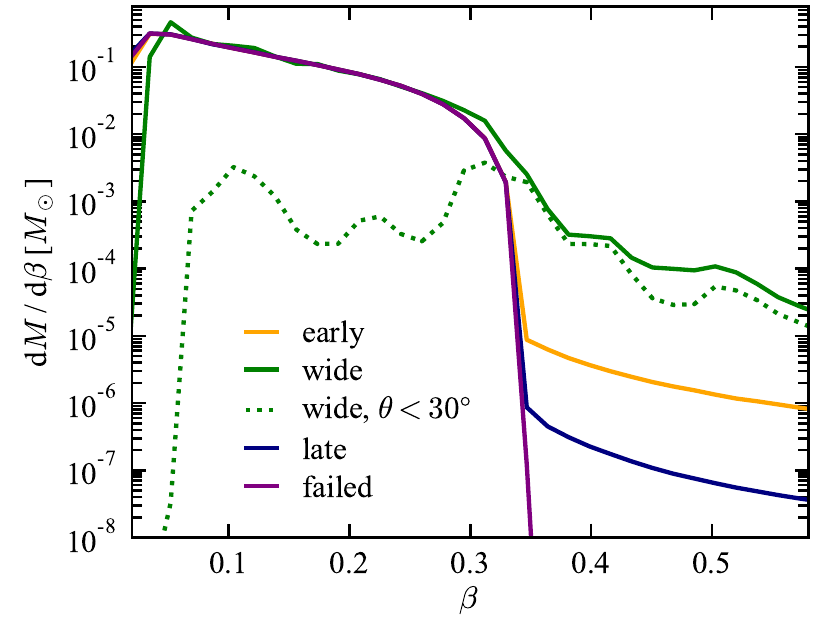}
  \caption{Distribution of ejecta mass $\mathrm{d} M / \mathrm{d} \beta$, as a function of velocity $\beta \equiv v/c$ for our ejecta models, denoted by different colors (orange: early; green: wide; blue: late; purple: failed). The dotted line only shows matter that lies within $30\dg$ of the pole. The solid lines show distributions for ejecta at all angles. In an angle-averaged sense, the primary differences between models is in the amount of high-velocity material. Near the pole, though, the jet can dramatically reduce the amount of low-velocity mass.}
    \label{fig:dmdb}
\end{figure}

We use equations \ref{eq:homol_rho} and \ref{eq:homol_total_therm} to evolve the hydrodynamics models from $100 \, \mathrm{s}$ to $900 \, \mathrm{s}$, when we begin the radiation transport calculation in Sedona. During this phase, we adopt the r-process radioactive heating rate from {figure 3 of} \citet{metzger:10}, assuming $\ye = 0.1$. Figure \ref{fig:2d_rho_T} shows the mass density $\rho$ and temperature $T$ of our starting models at 15 minutes. Note that for a given model, the density and temperature structures track each other closely.
High mass density leads to a high radioactive heating rate. The total thermal energy from radioactivity greatly exceeds the thermal energy of shock heating by the jet, so the density structure dictates the temperature structure.

We quantitatively compare the thermal energy from jet shock heating and radioactivity in figure \ref{fig:rad_heating}.
We show the total thermal energy as a function of time during the JET simulations. Due to the short duration of the jet-ejecta interaction, the jet heating is completely over by $\sim 10^{-5}\,\mathrm{days} \approx 1\,\mathrm{s}$.
Subsequently, the thermal energy decreases as $\propto t^{-1}$. 
Throughout, the energy from r-process heating exceeds that of the hottest jet-ejecta interaction model by nearly two orders of magnitude.

Figure \ref{fig:dmdb} shows the distribution of mass with radial velocity for the four models. Since we assume homologous expansion, the velocity distribution also corresponds to the spatial distribution of mass.  The jet does not interact with the majority of the mass of the ejecta. The angle-averaged mass distribution (solid lines) of low-velocity ($\beta \lesssim 0.35$) material is very similar across all models. {More energetic jets and those with wider opening angles impart more kinetic energy to the ejecta, leading to more high-velocity material.}

This high velocity material is confined to the polar regions. The dotted line shows the distribution of mass within $30\dg$ of the pole for the wide model.
At $\beta \gtrsim 0.35$, the mass distributions for the full ejecta and polar region only are very similar. At those same latitudes, the amount of low-velocity material is reduced by orders of magnitude. The vast majority of the contribution to low-velocity material comes from equatorial latitudes. Our assumption of homologous expansion at fixed velocity (equation \ref{eq:homol_rho}) neglects the fact that r-process heating per particle of $3\,\mathrm{MeV}$ can accelerate the gas up to $\sim 0.1c$. We may therefore somewhat underestimate the velocity of the low-speed part of the ejecta. {By assuming homology, we also implicitly assume that hot ejecta will not expand into the cavity opened by the jet. In nature, magnetic fields may keep the channel evacuated. The fate of the jet cavity remains an open question. Assessing it would require long-term ($>10\,\mathrm{s}$) magnetohydrodynamic simulations of neutron star merger ejecta.}

\subsection{Sedona}
\label{sec:sedona}

We use Sedona, a time-dependent, multi-wavelength, multi-dimensional Monte Carlo radiation transport code \citep{kasen:06, roth:15}, to determine the emission from the above models. 
The code tracks the emission of packets of radiant energy (``photons'') with a given wavelength.
Sedona calculates their propagation through the ejecta, accounting for random absorption and scattering events in the moving background. {All photon emission and scattering is done in the fluid frame. This accounts for relativistic effects such as Doppler shifting and beaming. It also ensures that the total energy of emitted photons is correct in the fluid frame.} the photons escape from the ejecta, they are tallied according to their escape time, wavelength, and propagation direction, giving time- and viewing-angle-dependent spectra and light curves. The ejecta are taken to expand homologously, such that the mass density in each cell falls off as $t^{-3}$. At each time step, the temperature of the cells is determined by equating the rate of thermal emission to the rate of radioactive heating plus photon absorption. Adiabatic losses emerge naturally from the scattering process.

The heating due to radioactive r-process products is not dominated by a single nuclide. Instead, there is an ensemble of relevant nuclides with a wide distribution of half lives \citep{metzger:10}. We do not track the decay and heating of all of these nuclides. We adopt the parametrized, time-dependent r-process heating rates of \citet{lippuner:15}, assuming an initial $\ye = 0.13$, expansion timescale $\tau = 0.84\, \mathrm{ms}$, and entropy $s = 32\,\kb$. We use the same parametrization throughout since the heating rate is largely independent of the microphysical parameters. Given the small impact of the jet on the thermal structure of the ejecta even on $\sim$ second timescales (Figure \ref{fig:rad_heating}), it is unlikely that the jet changes the composition of the bulk of the ejecta.

{The r-process heating rate, $\eta$, is calculated using the time in the lab frame. In doing so, we neglect time dilation in the fluid frame. Our heating rate approximates the power law, $\eta(t) \approx A t^{-1.3}$, so we underestimate the heating rate (energy deposited) by a factor of $\gamma^{1.3}$ ($\gamma^{0.3}$), where $\gamma$ is the fluid Lorentz factor. Our fastest material has $\gamma = 1.25$, so the r-process luminosity we calculate is incorrect by at most $30\%$. The effect on total energy deposited is even smaller, reaching only $5\%$. The omission of time dilation in $\eta$ will therefore have a small effect on our results.}

\subsection{Opacity}
\label{sec:methods_opacity}

Unless otherwise stated, we assume all material has a grey opacity of $1\cmspg$. This value is chosen to be roughly comparable to Planck mean line expansion opacities of first peak r-process elements (roughly, the second row of the $d$-block of the periodic table).
\citet{tanaka:20} find that at one day after ejection, these elements have mean opacities in the range of $\kappa \sim 10^{-2} - 10 \cmspg$. By contrast, Lanthanide-rich ejecta are expected to have mean opacities an order of magnitude greater \citep{kasen:13, tanaka:20}. In this study, we focus on the relative effect of different jet parameters on the light curves at varying viewing angles. As such, the exact choice of grey opacity does not affect our primary results.

To assess the possible impact of a more realistic, non-grey opacity, we re-run one of our models with temperature, density and frequency-dependent opacity. We include bound-bound, free-free, and electron scattering opacities. We use the line expansion formalism of \citet{karp:77} to tabulate bound-bound opacities on our discrete frequency grid.

For this calculation, we use an isotropic composition of half calcium-90 and half iron-90. Ideally, we would instead use a mixture of first- or second-peak r-process products. However, this is not possible due to our early start time and therefore high initial temperature. At the start of our calculations, the hottest portions of the ejecta are at a temperature of $5\times 10^6 \K$ and a density of $3 \times 10^{-5}\gcc$. Using the Saha equation and ionization energies from \citet{NIST_ASD} we can calculate the expected ionization state of a given element at that temperature and density.
At that point in $\rho-T$ space, we find that ruthenium (representative first-peak r-process product) is, on average, 41.7 times ionized, and that neodymium (representative second-peak/Lanthanide product) is 50.0 times ionized.
Therefore, in order to calculate the opacity for a realistic composition, we would need atomic levels and lines for over 40 ions of each element included. 
These are not currently available.
Opacities have been calculated out to temperatures of only $\sim 10^4 \K$ for the Lanthanides \citep{kasen:17, fontes:20, tanaka:20} and $\sim 10^5\K$ for d-block elements \citep{banerjee:20}. 

We use calcium and iron because they have similar atomic structure to first peak r-process elements, and therefore similar opacity patterns, at least for atoms that are up to ten times ionized \citep{banerjee:20}.
We also have atomic level and line data for all ionization states of these elements. To make as complete a line list as possible, we combine the atomic data of the Chianti \citep{dere:19} and CMFGEN \citep{hillier:01} atomic databases.
We use a mean molecular weight of 90, which is typical of first peak r-process products. When lines dominate the opacity, this choice more accurately reflects the actual number density of atoms in the ejecta. 
At the highest temperatures, though, this underestimates the opacity from electron scattering. The reason is that electron density from ionizing calcium and iron is artificially capped because they have fewer electrons than their counterparts one row below on the periodic table. 
The artificially lowered Thomson opacity will only affect the innermost parts of the ejecta at the earliest times. 
The photosphere is always at a low enough temperature (at the start, $2\times 10^5\K$ at a density of $1 \times 10^{-9}\gcc$) that the degree of ionization of first and second row d-block elements are comparable.

\section{Results}
\label{sec:results}

\begin{figure*}
  \includegraphics{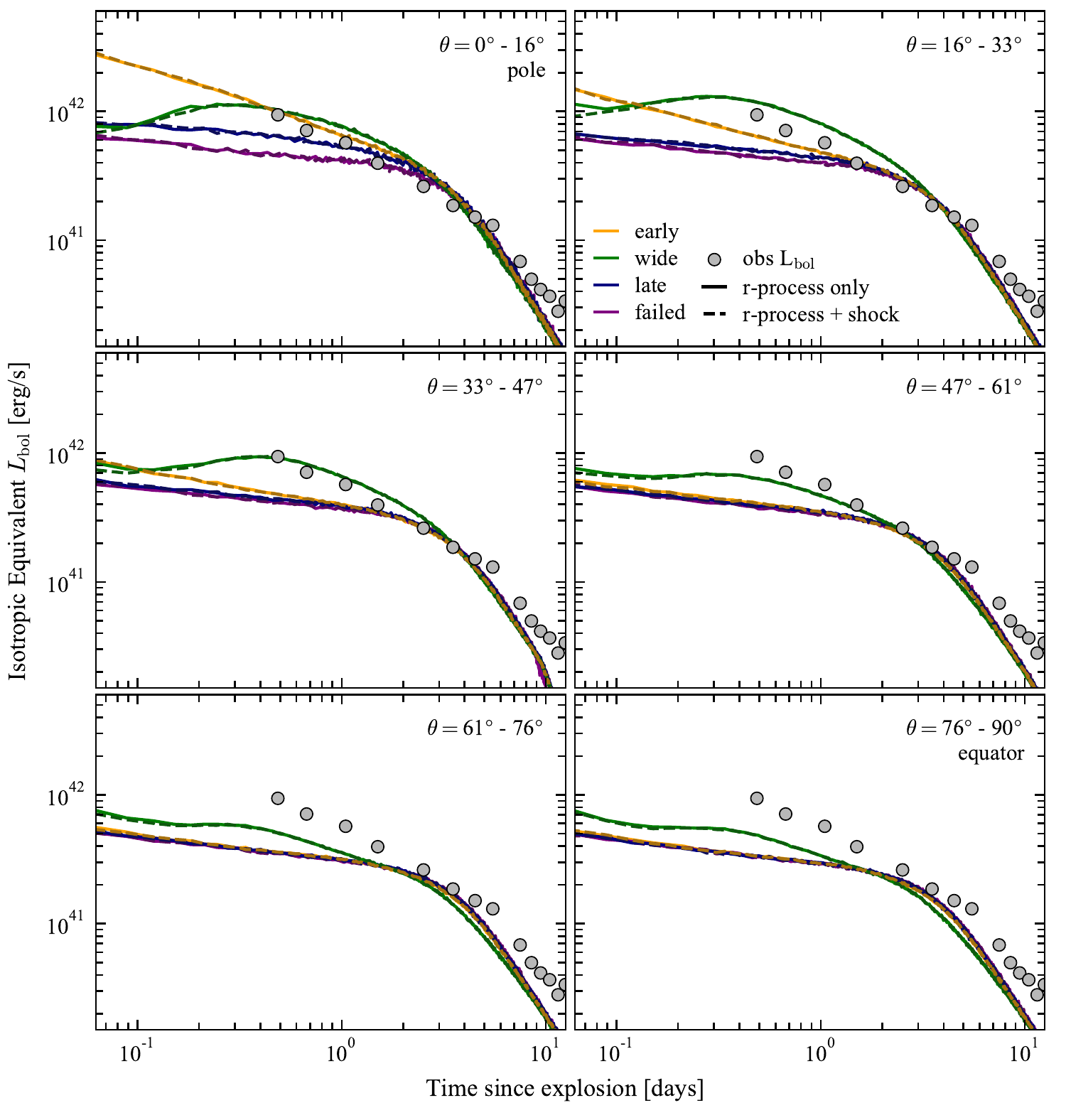}
  \caption{Isotropic equivalent bolometric luminosity at different viewing angles for different models. The bolometric luminosity of AT2017gfo \citep{drout:17} is overplotted in grey to guide the eye and demonstrate that our light curves are of approximately the correct luminosity and time scales. Each panel corresponds to a different viewing angle, arranged from polar to equatorial. Colors correspond to different jet-ejecta interaction cases, as described in Section \ref{sec:initial_models} and as shown in Figure \ref{fig:2d_rho_T} (orange: early; green: wide; blue: late; purple: failed). The light curves in solid lines consider only r-process heating, while the dashed lines also include jet shock heating. We find that the inclusion of jet shock heating does not appreciably affect the light curves. However, when the jet substantially alters the density structure of the ejecta, the polar light curves can be much brighter than on the equator. This is most pronounced in the early breakout case, where there is an energetic, collimated jet. By contrast, the model with a lower-energy jet and the same opening angle (late breakout) has a much less pronounced brightening on the pole.}
    \label{fig:lc_bol}
\end{figure*}

\subsection{Radioactive heating dominates over jet shock heating}
\label{sec:main_lc}

We simulate the viewing angle-dependent bolometric and broadband light curves for each of our four models. The bolometric light curves are shown in Figure \ref{fig:lc_bol}. Each panel shows a different viewing angle, and the colors indicate the different models (orange: early; green: wide; blue: late; purple: failed). To guide the eye, the observed bolometric luminosity of AT2017gfo constructed in \citet{drout:17} is overplotted in grey. We do not seek to exactly match the light curve of AT2017gfo, but do confirm that the time and energy scales in our models roughly match those of the event.

We find that the thermal energy imparted by the jet on the ejecta is negligible in comparison to the heating due to r-process decays. However, the kinetic energy of the jet is sufficient to change the density structure of the ejecta, leading to a viewing-angle variation in the radioactive transient.

The dashed lines in Figure \ref{fig:lc_bol} show light curves calculated considering the thermal energy from both jet shock heating and r-process heating.
The solid lines show curves from models that are identical to their dashed counterparts, other than that they include only r-process heating.
The light curves with and without jet shock heating are very similar, even $1\, \mathrm{hr}$ after merger. Since the jet shock heating in our models happens within the first several seconds of evolution, by $1\, \mathrm{hr}$ the contributions have largely adiabatically degraded.
Moreover, the thermal energy from shock heating is orders of magnitude less than that from r-process decays (Fig. \ref{fig:rad_heating}).

This is true even in the ``wide'' case, which has the largest jet energy by a factor of three and the largest solid opening angle by a factor of 16.
It also has the highest peak shock thermal energy, reaching $\sim 2 \times 10^{-2} \eej$. By contrast, the other models have peak thermal energies $< 10^{-3} \eej$ \citep{duffell:18}.
Even the large amount of thermal energy in the wide case is not sufficient to affect the light curve; radioactive heating dominates.

This demonstrates the difficulty of thermalizing sufficient energy for jet-ejecta shock heating to  directly impact the kilonova light curve. We find that {without a delay in jet onset,} any effects of the jet on the light curve after one to two hours will be due to changes to the density profile of the ejecta. {If the jet were delayed relative to the outflow, we would see more thermal energy from shock heating. Whether or not this would be sufficient to affect the light curve at one to two hours will depend on the length of the delay and is currently uncertain.} 

\subsection{Jet can strongly affect density structure}

The jet has a minimal effect on the density or energy structure of the failed model. We use the failed model to study the light curves from our ejecta model in the absence of a jet, allowing us to isolate the effects of the jet-induced asymmetry from the inherent viewing angle dependence of the light curves from the ejecta. This is particularly important since the initial ejecta distribution is slightly oblate (axis ratio $1.3$, \citep{duffell:18}).
The polar light curves in the failed case are $\sim 50$ per cent brighter than those on the equator due to the slightly larger surface area seen by an observer at $0^{\circ}$ \citep{darbha:20}.

The three other cases, however, show larger changes to the bolometric luminosity. The early and late cases show polar light curves that are brighter by factors of $\sim 3-4$ and $\sim 2$ respectively in the first $\sim 12 \,\mathrm{h}$. After this, the light curves lose most of their viewing angle dependence. In these models, the jet has carved out a low-density tunnel along its axis. This cavity can be clearly seen in the density distributions plotted in Figure \ref{fig:2d_rho_T}. These low density regions have very little radioactive heating relative to the unperturbed regions of the ejecta that remain much more dense.
The channel carved by the jet exposes the hot material that would otherwise remain optically thick until a few days after merger. The polar light curves are therefore brighter due to the hotter photosphere along the jet axis. While the jet cavity does slightly increase the projected surface area on the pole relative to the failed case, the effect is slight due to the small solid angle subtended by the jet. Instead, the primary cause of the increase in brightness is due to the higher photospheric temperature within the jet cavity.

\begin{figure*}
  \includegraphics{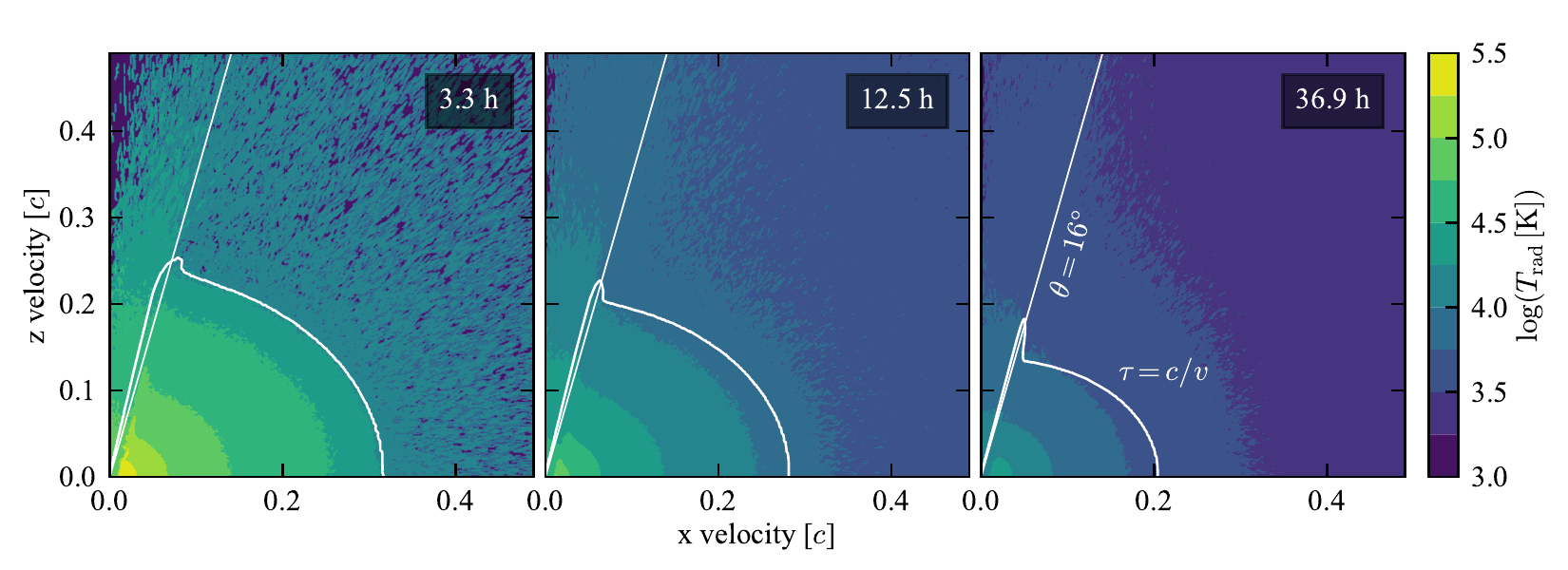}
  \caption{Slices of radiation temperature from radiation transport calculations of the early model at $3$, $12$, and $37$ hours after merger. The white contour marks where the radial optical depth $\tau = c/v$, where photons can efficiently diffuse out of the ejecta. This traces the density structure, showing the cavity carved by the jet. The hotter interior material becomes optically thin earlier than it would have in the absence of the jet.
  This leads to higher $T_\mathrm{rad}$ along the jet axis, giving brighter (Figure \ref{fig:lc_bol}) and bluer (Figures \ref{fig:spectrum_T_slice} and \ref{fig:lc_uvw2_r}) emission on the poles than elsewhere. The diagonal white line marks $\theta = 16^\circ$ from the pole, the maximum viewing angle included in the polar-most bin in our Sedona results.}
    \label{fig:T_slice}
\end{figure*} 

This effect is most easily seen in temperature snapshots from the radiation transport simulations. Slices of the radiation temperature at $3.3$, $12.5$, and $39.6$ hours are shown Figure \ref{fig:T_slice}. The white contour shows where the radial optical depth $\tau = c/v$. Temperatures at a given $\tau$ are higher at the poles than on the equator, and the ratio between polar and equatorial photospheric temperatures is greater at earlier times, dropping from 5 at 3.3 hours to 3 at a day and a half. Throughout, there is a drop in temperature at around 15 degrees, corresponding to the angle where the jet-induced asymmetry ends. Temperatures are also higher slightly away from the jet cavity (at say $\sim 20\dg$) than on the equator. 

\begin{figure}
  \includegraphics{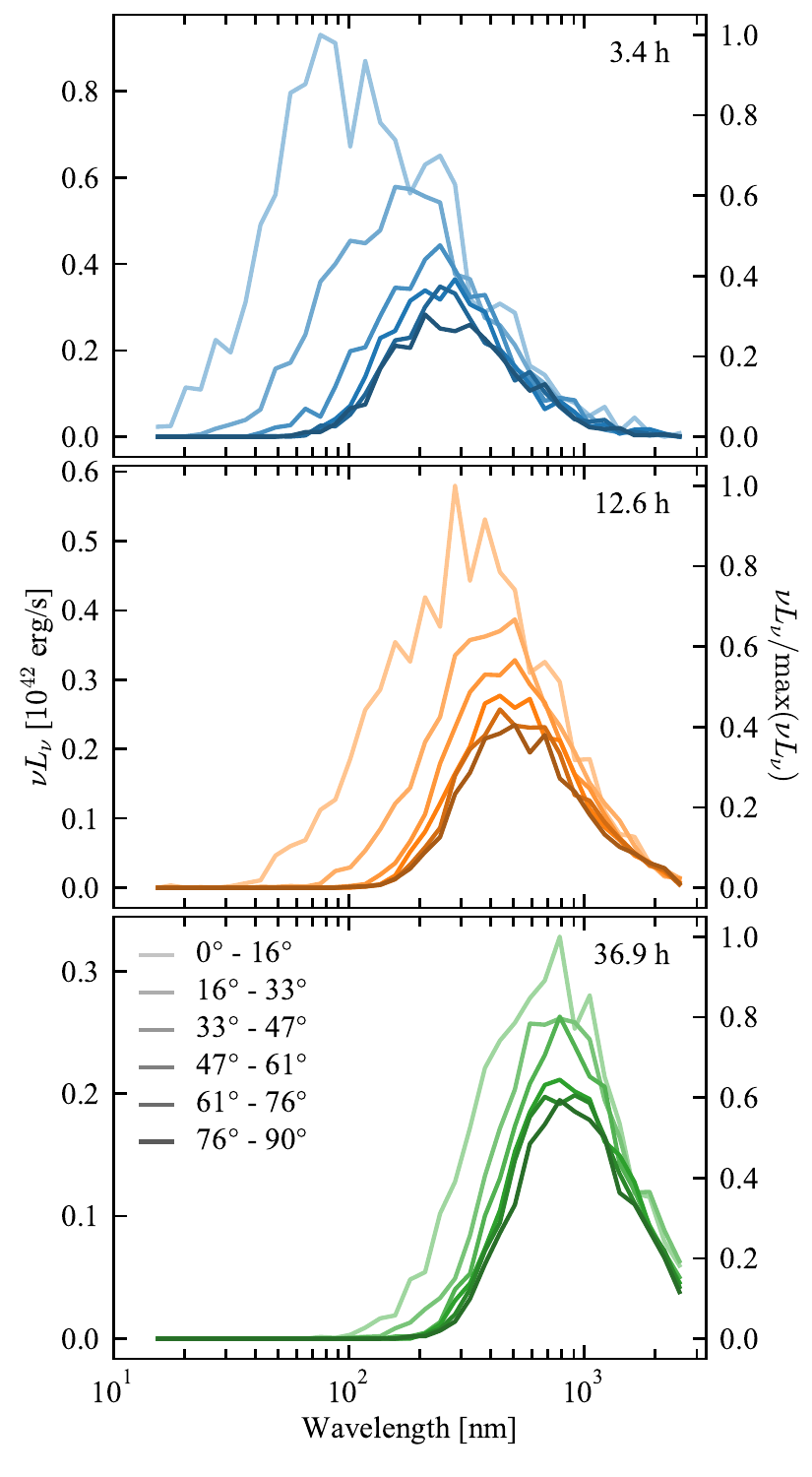}
  \caption{Viewing angle-dependent optical and UV spectra of the early breakout model. The times chosen approximately correspond to those shown in Figure \ref{fig:T_slice}. The left y-axis shows absolute $\nu L_\nu$, and the right axis is normalized to the maximum $\nu L_\nu$ across all spectra at that time. The viewing angles that are hottest in Figure \ref{fig:T_slice} are also substantially bluer in the spectra. At $\sim 3.4$ hours, the polar spectrum peaks at $\sim 90\, \mathrm{nm}$, while the equatorial one peaks at $\sim 300\, \mathrm{nm}$. The jaggedness of the spectra is due to statistical noise and does not represent lines. }
    \label{fig:spectrum_T_slice}
\end{figure}

The hotter photosphere on the poles leads to emission that is both brighter and bluer. This is apparent in the viewing angle-dependent spectra shown in Figure \ref{fig:spectrum_T_slice}. For the first day of evolution, the spectral energy distribution peaks brighter and at shorter wavelengths on the pole than on the equator. At $3.4\,\mathrm{h}$, the polar SED peaks at $\sim 100\,\mathrm{nm}$ as compared to $\sim 300\,\mathrm{nm}$ on the equator. Peak $\nu L_\nu$ is at around $1 \times 10^{42} \, \mathrm{erg/s}$ on the pole, as compared to $2 \times 10^{41} \, \mathrm{erg/s}$ on the equator. The polar spectra are wider than expected for a single-temperature black-body, indicating that the observer sees both the hotter polar region and cooler off-axis material. This highlights the need for multi-dimensional radiation transport simulations when predicting the light curves of non-spherical transients. At later times, the differences between polar and equatorial spectra become less pronounced, corresponding to the reduced temperature contrast between the two angles. Since the ejecta are also cooling, the spectra at all angles shift to redder wavelengths and are dimmer at later times.

\begin{figure*}
  \includegraphics{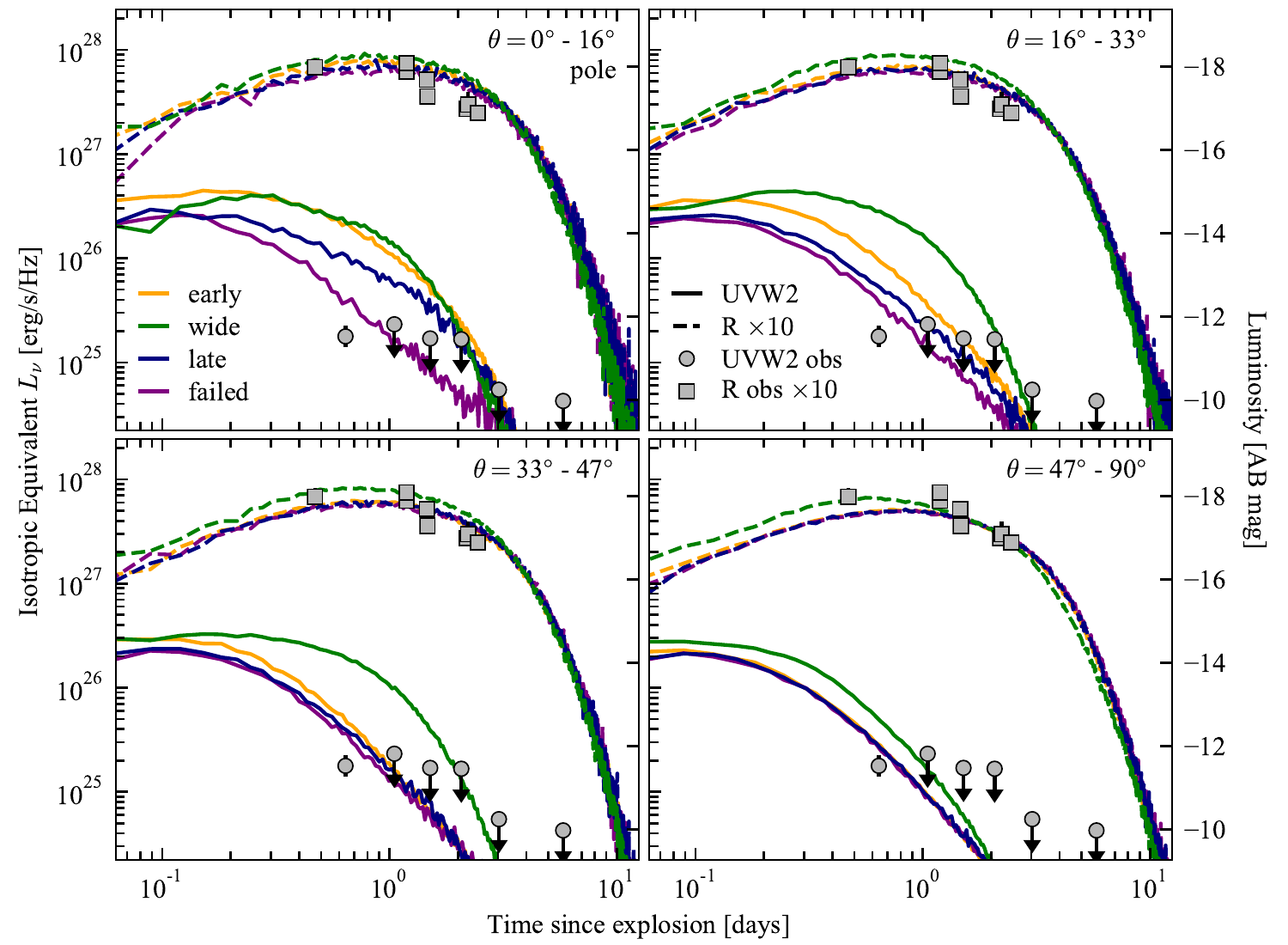}
  \caption{Isotropic equivalent luminosity in {\em Swift} in UVOT W2 (solid line) and Cousins R (dashed line, $\times 10$) bands at different viewing angles. These light curves are calculated in radiation transport calculations that account for both jet shock and radioactive heating of the ejecta. Different colors indicate different ejecta configurations, as described in Section \ref{sec:initial_models} and shown in Figure \ref{fig:2d_rho_T} (orange: early; green: wide; blue: late; purple: failed). Symbols show observed band luminosities of AT2017gfo \citep{kasliwal:17}. Arrows indicate upper limits. Jet-ejecta interaction disrupts the density structure near the poles, leading to higher polar photospheric temperatures, particularly at early times. Thus, the UV light curves are more strongly affected by the jet-induced asymmetry than those in R band.}
    \label{fig:lc_uvw2_r}
\end{figure*}

Bluer spectra at earlier times near the poles correspond to brighter ultraviolet light curves at those times. Figure \ref{fig:lc_uvw2_r} shows {\em Swift} UVOT W2-band (central $\nu = 193\, \mathrm{nm}$) and Cousins R-band (central $\nu = 635\, \mathrm{nm}$) light curves for our four models and compares them to observations \citep{kasliwal:17}. The UV light curves show a similar trend to the bolometric luminosities.
The early breakout model is brightest on the pole, and remains slightly brighter than the late and failed cases until about $45 \dg$ from the pole, where all models other than wide become indistinguishable. Since the wide case has the largest impact on the density structure, it also has the greatest effect on the light curves. At all viewing angles other than the pole, the wide breakout case is brighter than the other models in the UV. By contrast, the effect of the jet cavity on the R-band light curves is much smaller. The higher temperatures on the pole enhance emission across the UV and optical spectrum, but the enhancement is much less on the Rayleigh-Jeans tail of the black body distribution, where flux is proportional to temperature, rather than near the peak where the scaling is stronger.

\subsection{Comparison with afterglow}

\begin{figure}
  \includegraphics{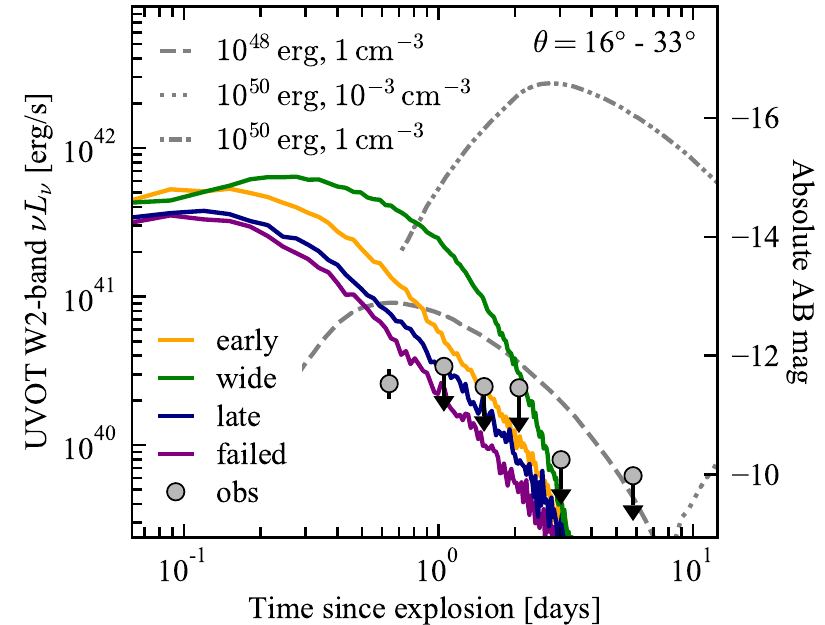}
  \caption{Isotropic equivalent luminosity compared to estimates of off-axis GRB optical afterglows. Colored lines show {\em Swift} UVOT W2 band light curves for different models just off-axis ($\theta = 16\dg - 33\dg$). Off-axis afterglow models from \citet{vaneerten:11} are overplotted in grey. We show estimates for combinations of different bipolar jet energies ($E \in \{10^{48}, 10^{50}\}\erg$) and ISM number densities ($n \in \{1, 10^{-3}\}\, \mathrm{cm}^{-3}$).}
    \label{fig:lc_uvw2_afterglow}
\end{figure}

Once the jet escapes the ejecta, it interacts with the surrounding interstellar medium, giving rise to synchrotron emission across the electromagnetic spectrum. Afterglows are brightest when viewed on-axis, and for energetic jets interacting with dense circumburst media \citep{sari:98, vaneerten:10, granot:18}. We expect the optical afterglow to outshine the kilonova if the system is observed pole-on. 
\citet{wu:19} fit the physical properties of the observed off-axis afterglow of GRB 170817A and use these parameters to predict what the event would have looked like on-axis. We convert their X-ray light curve to optical bands using the fitted spectral slope of the GRB 170817A afterglow, $L_\nu \propto \nu^{-0.6}$ \citep{margutti:18}. This predicts a UV luminosity of $-15\,\mathrm{mag}$ at one day, which is brighter than any of our on-axis kilonova models. This is consistent with predictions that, in general, an on-axis kilonova will be hidden from view by the GRB afterglow \citep{metzger:12}. In the redder bands, the kilonova will stay brighter for longer; it is possible that the kilonova will exceed the afterglow several days after the burst, at which point the early-time effects of the jet-induced asymmetry on the kilonova will have dissipated.

If observed off-axis, the kilonova may be brighter than the afterglow in UV bands for around one day. In Figure \ref{fig:lc_uvw2_afterglow} we compare our kilonova light curves with simulated off-axis afterglow light curves from \citet{vaneerten:11}. We consider combinations of circumburst number density $(n \in \{1, 10^{-3}\} \mathrm{cm}^{-3})$ and jet energy $(\ej \in \{10^{48}, 10^{50}\} \,\mathrm{erg})$. The densities are typical of short GRBs \citep{berger:14}, and the jet energies overlap with our more energetic models (early and wide). The brightest afterglow model we consider, ($10^{50}\,\mathrm{erg}, 1\,\mathrm{cm}^{-3}$), is likely to be an upper bound on the off-axis afterglow from one of our jet models. Even in that case, the UV kilonova will outshine the off-axis afterglow for the first half day of evolution.
    
\subsection{Non-grey opacities}
\begin{figure}
  \includegraphics{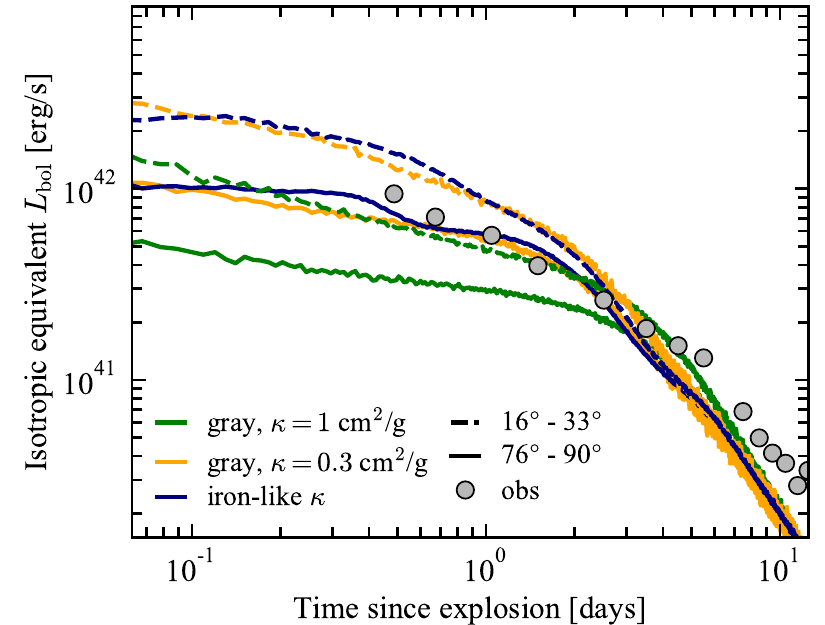}
  \caption{Isotropic equivalent bolometric luminosity at different viewing angles for the early breakout case. We compare iron-group-like opacities (blue, representative of first peak r-process elements), to bolometric luminosities for grey opacities of $1\cmspg$ (green) and $0.3\cmspg$ (orange). Solid lines show equatorial viewing angles, and dashed lines show a near-polar viewing angle. The bolometric luminosity of AT2017gfo is overplotted in grey circles \citep{drout:17}. All models have a similar viewing angle-dependence: they are brighter as seen from the pole than they are from the equator. The results for iron-group-like opacities are very similar to the grey $\kappa=0.3\cmspg$ results.}
    \label{fig:lc_bol_fe_opacity}
\end{figure}

All of the calculations that we have discussed thus far make the simplifying assumption that the ejecta have a constant grey opacity of $\kappa = 1\cmspg$.
This assumption is not realistic, since kilonova ejecta are expected to be rich in r-process elements, whose opacity is dominated by a dense forest of atomic lines.
There is therefore a concern that the brightening due to the jet-induced asymmetry could be obscured by atomic lines from d- or f-block elements that can blanket the blue and UV portions of the spectrum. 

To assess the effect of non-grey opacity, we re-simulate the ``early'' model but use a composition of half 90-Ca and half 90-Fe to calculate the opacity. We refer to this model as ``KFe''. This composition approximates the opacity of the first peak of the r-process (Section \ref{sec:methods_opacity}). 
The isotropic equivalent bolometric luminosity is shown in Figure \ref{fig:lc_bol_fe_opacity}. We compare these light curves with those of models with grey opacity $\kappa = \{0.3, 1\}  \cmspg$ (K0.3 and K1.0, respectively). The figure shows the light curves seen by near-polar and equatorial observers. We do not show a pole-on kilonova light curve, since at those angles, it would be much dimmer than the accompanying afterglow. The bolometric light curves of the KFe model roughly match those from K0.3; both KFe and K0.3 are brighter than K1.0, but have the same shape. The viewing angle-dependence remains in all models; near-polar viewing angles are brighter than their equatorial counterparts.
\begin{figure}
  \includegraphics{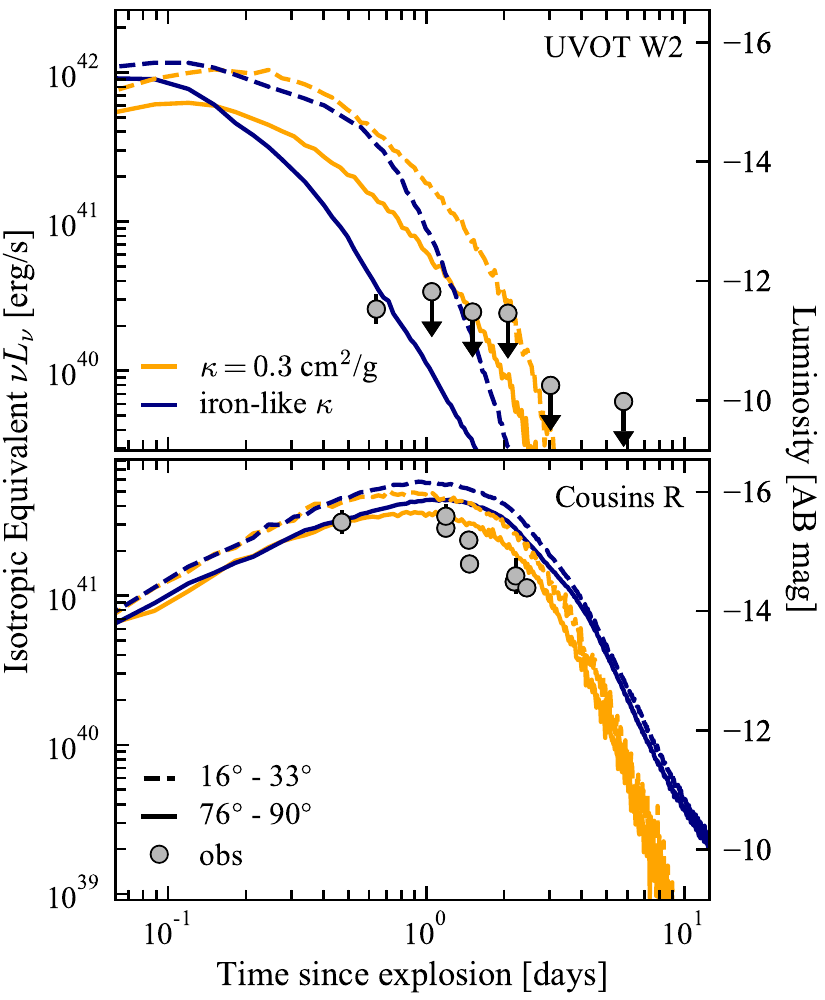}
  \caption{Isotropic equivalent band luminosity (\emph{Swift} UVOT W2 and Cousins R) at different viewing angles for iron-like opacity (blue), which we take to be roughly representative of the first r-process peak. We compare these light curves with those from a model with a grey opacity of $0.3\cmspg$ (orange). Observations of AT2017gfo are shown in grey circles; arrows indicate upper limits \citep{kasliwal:17}. We show light curves as seen from the equator (solid) and near the pole (dashed). The UV light curve falls more rapidly with more realistic opacities than the grey model, particularly on the equator. This is due to falling temperatures, which cause the iron and calcium to recombine, increasing the bound-bound opacity. The R-band light curve is largely unaffected by the choice of opacity.} 
    \label{fig:lc_UVW2_r_fe_opacity}
\end{figure}

While bolometric light curves of grey (K0.3) and more realistic (KFe) models are comparable, the band luminosities differ, particularly in the UV. Figure \ref{fig:lc_UVW2_r_fe_opacity} compares \emph{Swift} UVOT W2 and Cousins R band light curves for the KFe and K0.3 models. The primary effect of a more realistic opacity is that the UV light curve evolves more quickly, peaking earlier and brighter and dimming more rapidly. This is most pronounced near the equator, where the KFe model peaks at $-15.4$ mag, and dims by $4.8$ mag within the first day of evolution. By contrast, at the same viewing angle, K0.3 peaks slightly dimmer at $-15.0$ mag but only dims by 2.5 mag by the end of the first day.

The dimming in the UV is due to the rise in bound-bound opacity at ultraviolet wavelengths as the ejecta cool and recombine. Consequently, this dimming is less pronounced near the pole, where the temperature is higher. As seen from near the pole, KFe looks very similar to K0.3 for the first $\sim 20 \,\mathrm{h}$, before there is a sharp turnover in the KFe UV light curve that is not matched by K0.3. 

When UV absorption increases, there is a corresponding increase in re-emission across the spectrum. Photons are effectively scattered from high to low frequency. This leads to a slight enhancement to the R-band light curve after about a day. Overall the R-band light curve is largely unaffected by the choice between iron-like and comparable grey opacity, reflecting the low bound-bound opacity in the infrared and red bands.

\section{Discussion \& Conclusion}
\label{sec:discussion}
A relativistic jet in binary neutron star or neutron star-black hole mergers can affect the light curve seen from a homologously expanding, radioactive outflow. These changes are due to the impact of the jet on the density structure of the ejecta. The jet evacuates a region around its axis, which exposes the hotter inner material that would otherwise remain optically thick. The photosphere in those areas is hotter, and the emission from near the pole is brighter and bluer. The bolometric (ultraviolet) light curves can be enhanced by factors of a few (several) at near-polar viewing angles on $\sim\mathrm{day}$ timescales (see first panels of figures \ref{fig:lc_bol} and \ref{fig:lc_uvw2_r}). The effect is less pronounced $15\dg - 30\dg$ from the pole, though still noticeable. Past $30\dg$ and in (infra)red bands, light curves are largely unchanged. It is unlikely that an off-axis afterglow would outshine the kilonova and hide this effect (figure \ref{fig:lc_uvw2_afterglow}). The degree of brightening depends on the jet parameters. Jets with larger energies and opening angles create larger cavities within the ejecta, so the light curves are more strongly affected. The effects are also not as confined to polar viewing angles (see figures \ref{fig:lc_bol} and \ref{fig:lc_uvw2_r}).

Our approximate treatments of both the homologous outflow structure and the opacity of the ejecta introduce uncertainty into our quantitative predictions. However, the qualitative difference between the polar and equatorial light curves of a given model and the rough magnitude of the effect should be robust, {provided that a cavity exists}. 
We have shown that an evacuated region within an otherwise spheroidal outflow should lead to brighter and bluer emission from inside the cavity. This result should generalize to other similar ejecta configurations, such as magnetically-driven outflows from NS merger accretion disks.

{The hydrodynamic simulations that underlie our work are axisymmetric and ignore magnetic fields, which may result in an artificially underdense cavity. In 3D magnetohydrodynamics (MHD), jets may develop kink instabilities, which can disrupt their propagation \citep{bromberg:19}. There may also be mixing between the jet and the surrounding cocoon \citep{gottlieb:20}. Both of these effects could increase the density within the jet cavity and reduce the polar brightening. There is also a question of whether hot ejecta will expand latitudinally into the jet cavity, reducing its size or erasing it entirely. Magnetic pressure within the jet cavity may stabilize the channel, but the circumstances under which the cavity is stable are uncertain. Further study of the long-term ($>10\,\mathrm{s}$), 3D MHD evolution of neutron star merger ejecta is needed to better understand the jet cavities that may form and persist in nature.}

Because the impact of the jet cavity is largest in the blue and ultraviolet (see figures \ref{fig:spectrum_T_slice} and \ref{fig:lc_uvw2_r}), our results point to the importance of early UV followup of gravitational wave signals and highlight the value of wide-field UV surveys.

This picture is largely insensitive to our choice of opacity. Emission is brighter and bluer near the pole than on the equator for models that have a small grey opacity as well as for models that have an opacity representative of first peak r-process elements (see figures \ref{fig:lc_bol_fe_opacity} and \ref{fig:lc_UVW2_r_fe_opacity}). 
The UV light curve in this model falls off more quickly than in the  grey case. As the ejecta cool and recombine, the bound-bound opacity increases in the blue and ultraviolet. The ejecta are hotter on the pole than on the equator, so the UV light curve stays bright for about a day near the pole, as compared to a few hours further off-axis. {We also expect our qualitative results to be independent of any latitudinal variation in opacity, since the effects of a jet cavity on light curves are largely confined to near-polar viewing angles. Material near the poles is expected to have lower rather than higher opacity \citep{miller:19_diskblue}. Polar light curves may generally be brighter and bluer, even without a jet due to the lower opacity and larger projected surface area \citep{korobkin:20}. A jet cavity would enhance this effect.}

Our results complement recent work that has studied the effect of a jet on a layer of high-opacity material that may form in a neutrino-driven wind from the NS merger accretion disk \citep{nativi:21}. Disrupting this layer of occluding material brightens the early kilonova as seen from near the pole. The effects of a jet on the long-term evolution of NS merger ejecta, and therefore on kilonova light curves, remains a rich and interesting direction for future work.

While some of the jet energy is thermalized in the ejecta via shocks, this contribution to the total thermal energy of the ejecta is negligible compared to that from radioactive heating at all of the times we study $(\gtrsim 1\,\mathrm{h})$. 
 These results are consistent with those of \citet{gottlieb:20}, who find that cooling emission from jet shock heating becomes subdominant by $1\,\mathrm{h}$.

Our models assume a minimal delay between the ejection of the more spherical outflow and the start of the jet engine. In our highest jet energy per solid angle case (``early'' in table \ref{tab:models}), the jet breaks through the ejecta at a radius of $\sim 10^7\,\mathrm{cm}$. In other cases we consider (``late'' and ``wide'' in table \ref{tab:models}) the jet/shock breaks out at $\sim 10^{10}\, \mathrm{cm}$, after about a second of expansion. After breakout, the deposited thermal energy falls off as the reciprocal of the radius of the ejecta. By delaying the start of the engine and using a wide jet opening angle, some prior work has suggested that the cooling emission from shock-heated regions can reach luminosities in excess of $10^{42}\, \mathrm{erg/s}$ on timescales of several hours to days \citep{kasliwal:17, piro:18}. However, the energies supplied by jet heating in our calculations --- which have {jet opening angles and energies typical of short GRBs \citep{fong:15}}--- are factors of $100-1000$ less than that supplied by the r-process (figure \ref{fig:rad_heating}). In order for shock heating to be significant for the kilonova light curve, the jet energy would need to be increased by a large factor $\sim 100$, the ``jet'' would need to be relatively spherical, and/or there would need to be shocks between the jet and ejecta at distances much greater than the $10^{10}\,\mathrm{cm}$ found here. We regard these as unlikely.

\section*{Acknowledgements}
This research was funded by the Gordon and Betty Moore Foundation through grant GBMF5076. HK was supported in part by a DOE Computational Science Graduate Fellowship under grant number DE-FG02-97ER25308. EQ and DK were also supported in part by Simons Investigator awards from the Simons Foundation.The simulations presented here were carried out and
processed using the National Energy Research Scientific Computing Center, a U.S. Department of Energy Office of Science User Facility, and the Savio computational cluster resource provided
by the Berkeley Research Computing program at the University of
California, Berkeley (supported by the UC Berkeley Chancellor,
Vice Chancellor of Research, and Office of the CIO).

\emph{Software:} Sedona \citep{kasen:06, roth:15}, \texttt{matplotlib} \citep{matplotlib}, \texttt{numpy} \citep{numpy},  \texttt{scipy} \citep{scipy}, FSPS filter files \citep{conroy:09, conroy:10}.

\section*{Data Availability}
The data underlying this article will be shared on reasonable request to the corresponding author.




\bibliographystyle{mnras}
\bibliography{bibliography,non_ads} 



\appendix




\bsp	
\label{lastpage}
\end{document}